# The crossed-sine wavefront sensor: first tests and results

Laura Schreiber*[a], Yan Feng[b], Alain Spang[c], François Hénault[b], Jean-Jacques Correia[b], Eric Stadler[b], David Mouillet[b]

[a]INAF, Osservatorio di Astrofisica e Scienza dello Spazio di Bologna, Via Gobetti 93/3, 40129 Bologna, Italy ; [b]Institut de Planétologie et d'Astrophysique de Grenoble, Université Grenoble-Alpes, CNRS, B.P. 53, 38041 Grenoble, France ; [c]Université Côte d'Azur, Observatoire de la Côte d'Azur, CNRS, Laboratoire Lagrange, France

**ABSTRACT**

The crossed-sine wavefront sensor (WFS) is a pupil plane wavefront sensor that measures the first derivatives of the wavefront. The crossed-sine WFS achieves a simultaneous high spatial resolution at the pupil of the tested optics and absolute measurement accuracy comparable to that attained by laser-interferometers, but with a much more compact, cheaper set-up, compatible with polychromatic light. It is made by three main components: a gradient transmission filter (GTF) built from a product of sine functions rotated by 45 degrees around the optical axis, a 2x2 mini-lens array (MLA) at the focus of the tested optical system and a detector array located on a plane conjugated to the pupil. The basic principle consists in acquiring four pupil images simultaneously, each image being observed from different points located behind the GTF. After the simulation work which demonstrated the wavefront reconstruction capability, we are now in the phase of implementation of the prototype in the lab. In this paper we introduce seven customized phase masks and make measurements of them. First tests and results are demonstrated, based on which we explore the performance of our crossed-sine WFS and make comparisons with that of the laser-interferometer.

## 1. INTRODUCTION

Wavefront sensors have now become core components in the fields of adaptive optics for astronomy[1], biomedical optics[2], and metrology of optical systems[3]. However, it remains challenging to achieve simultaneously high spatial resolution at the pupil of the tested optics and absolute measurement accuracy comparable to that attained by laser-interferometers. We have recently demonstrated[4] a new WFS concept, the crossed-sine WFS[5][6], which could achieve both previous goals maintaining a simple design. In French, the project of this crossed-sine WFS is called "Analyseur de Surface d'Onde de Nouvelle Génération (ASONG)" which means "new generation wavefront sensor". Hence ASONG will be used to define the crossed-sine WFS in the following text.

The general measurement configuration of ASONG is illustrated in Figure 1 which consists of the following essential elements:

- An optical measurement head achieving the simultaneous acquisition of four grayscale pupil images;
- A computation unit for the digital reconstruction of the wavefront error (WFE) from the four acquired images;
- The optical system to be tested;
- A light source or an illuminated object, emitting or reflecting beams through the tested optical system up to the measurement head.

*laura.schreiber@inaf.it; https://www.linksium.fr/en/projects/asong

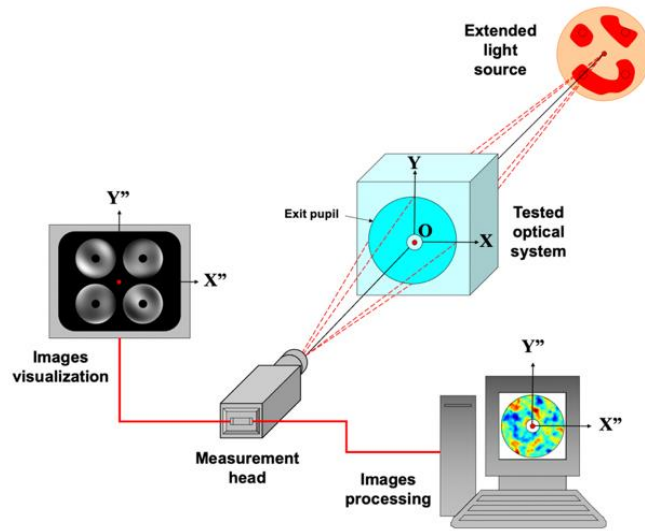

Figure 1. General measurement configuration of ASONG.

The design of key optical elements of the ASONG system, i.e. gradient transmission filter (GTF) and mini-lens array (MLA), including the general principles and fabrication requirements, has been illustrated in our previous publications[4][5]. In this paper, we present the preliminary test results to characterize the first complete version of the ASONG laboratory prototype. For the sensor characterization we rely mostly on i) low order (defocus) WFE measurements by introducing a known shift at the light source level, and ii) by the means of seven customized high-order phase masks measured with a laser-interferometer.

## 2. THE ASONG FIRST PROTOTYPE

The first ASONG wavefront sensor prototype is shown in Figure 2. Its physical size fits a volume of 92 x 52 x 40 mm$^3$, including the alignment tool. As a reference for the expected volume of the final version of ASONG, the implemented camera has a side of 30 mm. The mechanical parts are anodized in order to avoid internal unwanted reflection. The main optical components (GTF and MLA) box can be adjusted in focus, centering and rotation.

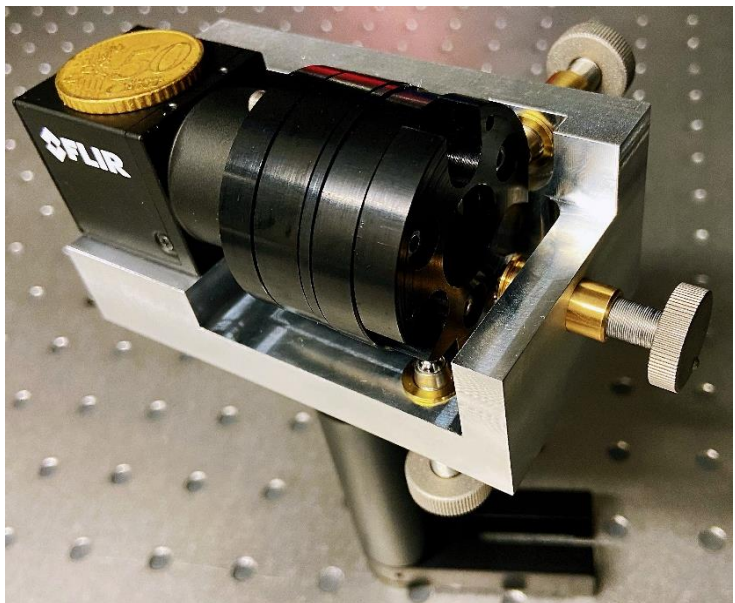

Figure 2. Picture of the first realized ASONG prototype including alignment tool

The value of the period of the GTF implemented in the actual prototype assures both large sensitivity and dynamic range. Different technologies and materials are under investigation for the GTF and MLA fabrication. The actual prototype mounts a 2 mm thick glass GTF and a PMMA MLA. A better pupil optical quality, and consequently higher performance, is achievable using glass MLA. Another view of the prototype is shown in Figure 3, where the optical key components of the wavefront sensor (GTF on the left panel and MLA on the right one) are shown as an integrated part of the prototype.

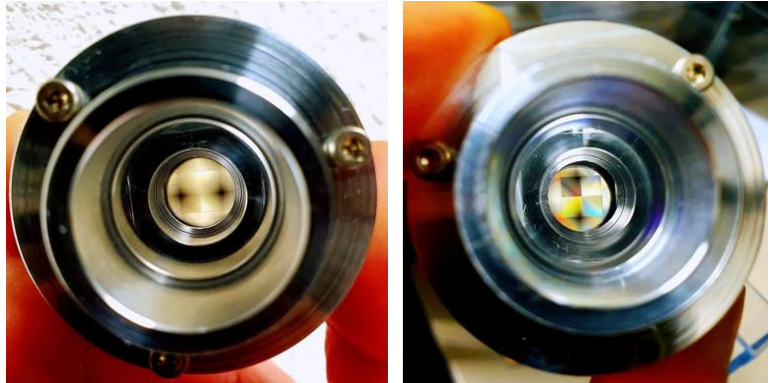

Figure 3. The ASONG optical key components integrated in the prototype

## 3. TEST OPTICAL CONFIGURATION

A preliminary prototype was procured and integrated to assess the ASONG wavefront sensing capability in terms of resolution and precision. The test set-up is shown schematically in Figure 4.

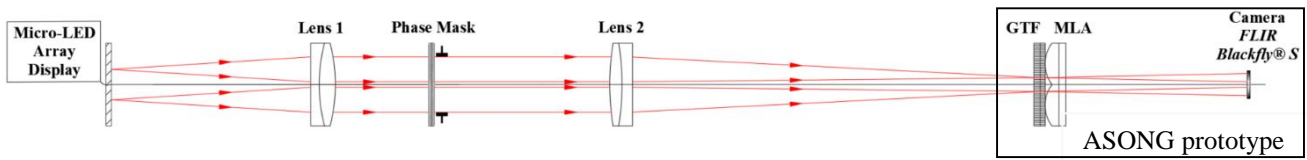

Figure 4. The optical configuration of the ASONG prototype and its test set-up.

A micro-LED array display (Jade Bird Display JBD5UM720P-R that yields 720p resolution with 633 nm wavelength and 1,9μm diameter emitters on 5μm pitch) is applied for creating "free-pattern" and possibly extended light sources. An achromat "Lens 1" (focal length $f_1$ 150 mm) collimates the light. The collimated beam goes through a pupil mask with a pupil of 25 mm in diameter, and is refocused by a second achromat "Lens 2" (focal length $f_2$ 500 mm). The ASONG prototype described in Section 2 is placed in order to have the GTF on the focal plane of compound Lens 1&2 system, while the MLA is consequently ~ 0.5 mm away. The prototype detector records four well-sampled (1000 x 1000 pixels) images of the pupil with a pixel size of 2.4 μm (FLIR Blackfly S). An 'in house' pipeline written in IDL® (Interactive Data Language) is used to compute the wavefront derivatives from the intensities variation of the pupil images as described in [7] and to deliver WFE measurements.

In order to test and characterize the WFS, the ideal configuration is to add to the system a known WFE and to compare it with the one measured by our prototype. As illustrated later on in this paper, two approaches have been used to introduce known aberrations:

1) Known defocus and tip-tilt WFE can be easily introduced by acting on the source position w.r.t. the Lens 1 (or on the prototype position w.r.t. the Lens 2). For this reason, the Micro-LED array has been mounted on X-Y linear stages with a 50 mm range and 10 μm steps;

2) Phase masks with specific patterns can be positioned in the set-up pupil plane and measured with the WFS. In this case a reference measurement of the phase mask, taken with a competitor instrument like a laser-interferometer, is mandatory.

## 3.1 The phase mask design and optical configuration

We have customized a phase plate to introduce different kinds of WFE. The phase plate is made of a sandwich of CNC machined acrylic and cast optical polymer by Lexitek (US). The surface is machined with the design optical path difference (OPD) scaled by $1/\Delta n$ (being n the refraction index). Seven different phase masks were machined in the support following the geometry of Figure 5(a).

The mask patterns have different characteristics (rms, PtV) and different combinations of high and low special frequencies. They are respectively named as 'ripple' (polishing errors), 'sinus' (periodic errors given by the combination of a few periodic functions with different periods), 'defocus' (Low order Zernike 2), 'zer55' (medium order Zernike 55), 'eye' (typical eye aberrations recovered by literature[8]), 'Kolmogorov' (typical aberration introduced by the earth atmosphere), and 'residual' (residual atmospheric aberration after adaptive optics loop correction of the first 200 modes, for the astronomical application). Each of the circular masks has a diameter of 25 mm. The WFE RMS of the phase masks ranges approximately between 0.1 and 1 micron, corresponding to PTV approximately from 1 to 6 microns. The phase plate support is 100 mm diameter with 83 mm of active area (22 mm thickness). Figure 5(b) shows the motorized stage for the correct positioning of the chosen phase mask in the optical path.

Table 1: Summary of phase masks characteristics

| Name | RMS (micron) | PTV (micron) | Description |
|---|---|---|---|
| Ripple | 0,18 | 0,88 | polishing errors |
| Sinus | 0,24 | 1,01 | periodic errors |
| defocus | 0,63 | 2,18 | defocus |
| zer55 | 0,15 | 1,35 | zernike 55 |
| eye | 0,45 | 2,57 | eye aberration |
| kolmo | 1,10 | 6,03 | Atmosphere |
| residual | 0,17 | 1,42 | First 200 modes corrected |

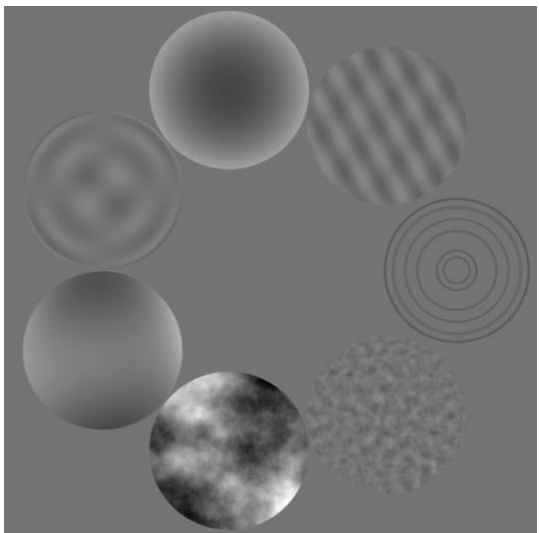
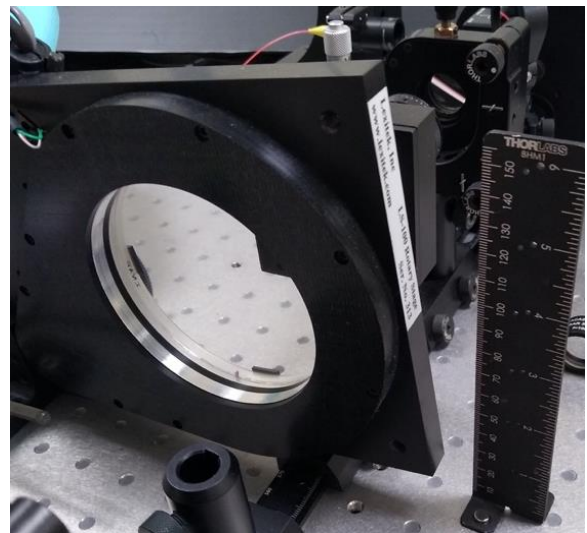

(a)         (b)

Figure 5. The phase plate: (a) the designed patterns; (b) the motorized mount.

## 4. PERFORMED TESTS AND RESULTS

After having proved the general ASONG concept in previous works[5], we concentrated on the characterization of the WFS. In particular we measured the sensitivity, the linear range and the WFE measurement accuracy for different configurations of the WFS. We were not able to measure the entire dynamic range of the system due to limits in the linear stage maximum range and to limits in phase masks manufacturing (maximum slope around 5 milliradians). For a preliminary estimation of this specification we have extrapolated and reported in the conclusive table a value for which we feel confident (as from the design analysis and early experimental results), but further measurements are planned for the near future. All the measurements have been done in high SNR (Signal to noise ratio) conditions and in monochromatic light (LED 0.633 μm).

### 4.1 Accuracy estimation

Accuracy is defined as how close or far off a given set of measurements are to their true value. In order to have a fair estimation of the accuracy, it is mandatory to have, as a reference, an accurate measurement of the introduced WFE. We decided to perform a very simple experiment to estimate the measuring accuracy of our prototype independently to any other measurement device since we estimated that all the available measurement devices, including interferometers, should perform an accuracy comparable to the ASONG one. Furthermore, part of the measurement discrepancy could be due to set-up aberrations and measurement matching software techniques. We therefore added a known defocus amount (WFE rms) by shifting the source of a known amount. The theoretical values can be then compared to the measured ones point by point and an average value of accuracy can be estimated as the average distance of the measurements from the theoretical defocus line (see Figure 6). This value is 0.005 μm, that corresponds to about $\lambda/130$ ($\lambda = 0.633$ μm). This number is of the same order of the defocus introduced by the translation stage positioning sensitivity. We then speculate that this value represents a conservative estimation. We recall that the first ASONG prototype is built with first attempt key optical components and higher quality GTF and MLA are available for testing since the measurements were made. The results will be reported in future publications. Higher accuracy measurements, using a different set-up emulating the GTF through an SLM and using a glass optical component to re-image the pupil on the camera, where described in a previous paper[4] and visible in    Figure 7.

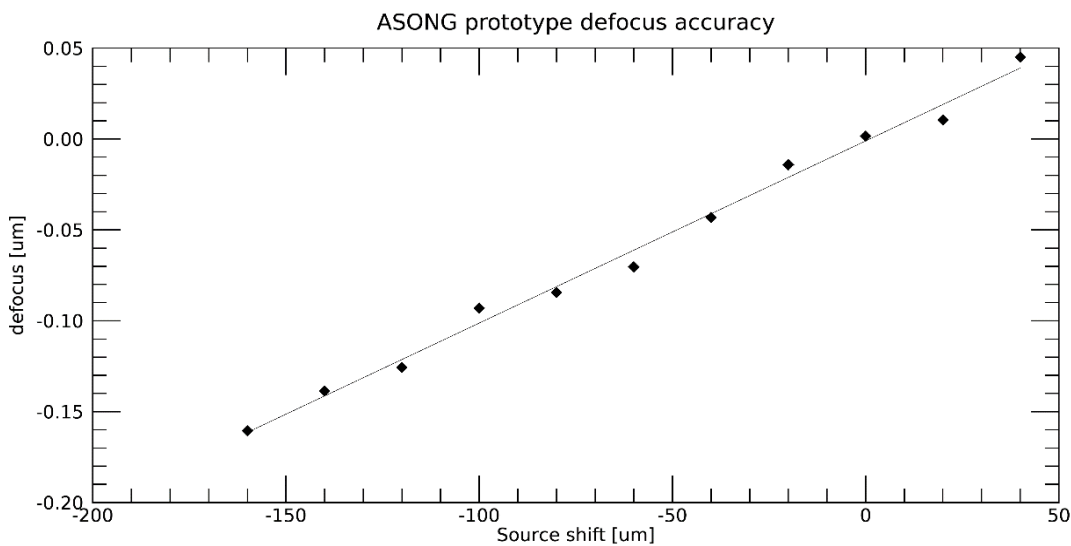

Figure 6. ASONG prototype accuracy measurement. Line: true defocus values; diamonds: ASONG measurements

### 4.2 Sensitivity estimation

Sensitivity is defined by the smallest WFE detectable by the WFS. A simple way to estimate it is by introducing a small WFE and check whatever the WFS is able to measure it and with which accuracy and precision. We did it by introducing a small defocus to the system again by shifting the source of a known amount. The rms WFE from defocus can be written as[9]:

$$\sigma_{def} = \frac{1}{16\sqrt{3}} \frac{D^2}{f_1^2} \delta f_1 \qquad (1)$$

where $D$ is the pupil diameter and $\delta f_1$ represents the source shift w.r.t. Lens 1. A first estimation of the sensitivity was made using the set-up described in [4], where an SLM was implemented to mimic the GTF and glass optics were used instead of the PMMA MLA integrated in the first prototype. The GTF period of this preliminary prototype was the same (9.05mm), while the pupil size was smaller (9 mm diameter) with the same pupil sampling (1000 x 1000), that translates into an higher resolution. This higher resolution should not impact the measurement of a low order like the defocus, making this measurement valid for our purposes. The achromat "Lens 1" has the same focal length f1 of 150 mm. The source was mounted on a translation stage with a minimum step of 10 μm. Putting these numbers in equation (1) we find that this shift corresponds to a WFE rms $\sigma_{def}$ of 1.3 nm that correspond to about λ/500. From Figure 7 it is apparent that this WFE increment is well detectable, with a very high accuracy.

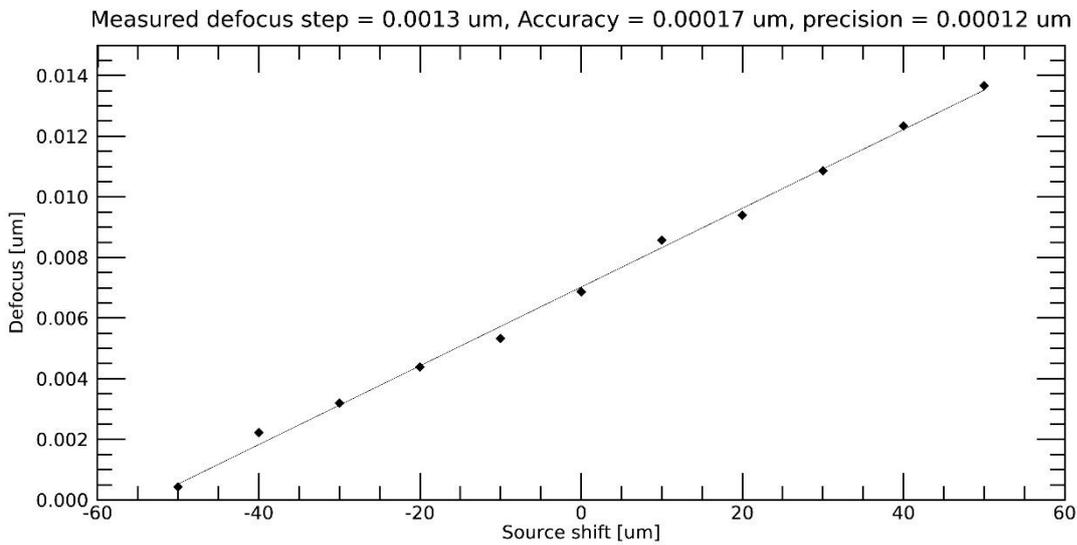

Figure 7. Sensitivity measurement indicates that a WFE increment of about λ/500 is well detectable

### 4.3 Linear range estimation

One of the advantages of ASONG comes from its linear response on a wide range of WFEs. As already shown in Section 4.2, ASONG, in condition of high SNR, is sensitive to very small WFE (centesimal fractions of lambdas). The WFS response remains constant on a very wide range of WFEs, of the order of tens of lambda PtV, preserving a very high absolute accuracy. Figure 8 shows the same plot as represented in Figure 6 but extended on a wider range of defocus WFE reported in lambdas PtV for the sake of clarity. It is remarkable that, even at the extremes of the curve, the pupil images reveal that the WFS is still far from the saturation (Figure 9). The features visible in the pupil image are due to the optical quality of the PMMA MLA and by the dust and they start to be visible when big defocus signal is present. Deeper studies of the entire linear range of the sensor is scheduled for the near future.

The latest prototype sensor enabled us to fully measure the wavefront of various phase mask (see next chapter) and, as said before, there is technological limit during their manufacturing process that allows only to engrave slopes up to 5 milliradians. This later value, if taken at the edge of the pupil of the same prototype, converts roughly to a estimated +/- 50 lambda PtV defocus. Since the linearity range has not yet been characterized that far, deeper studies are scheduled for the near future.

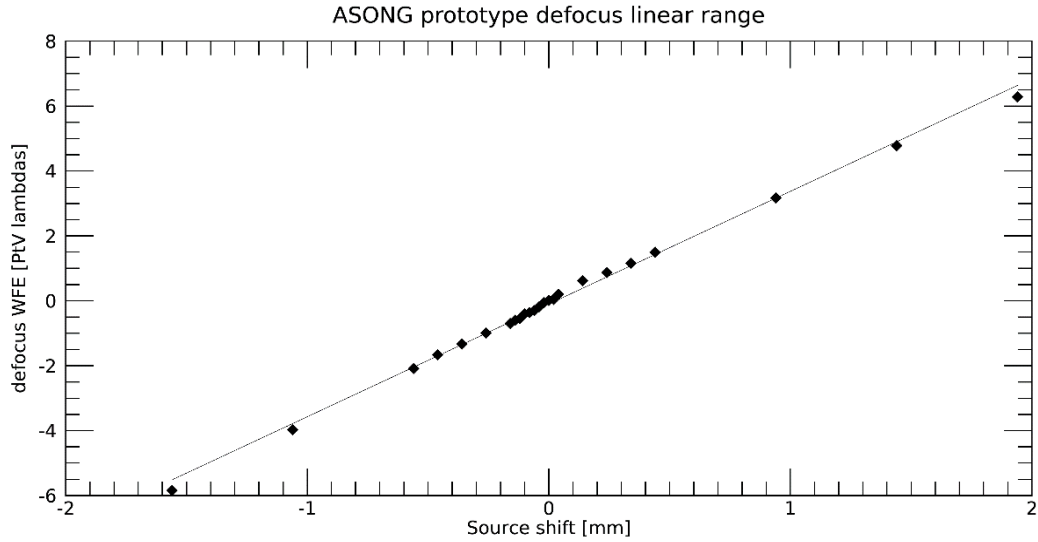

Figure 8. ASONG Defocus linear range PtV in lambdas

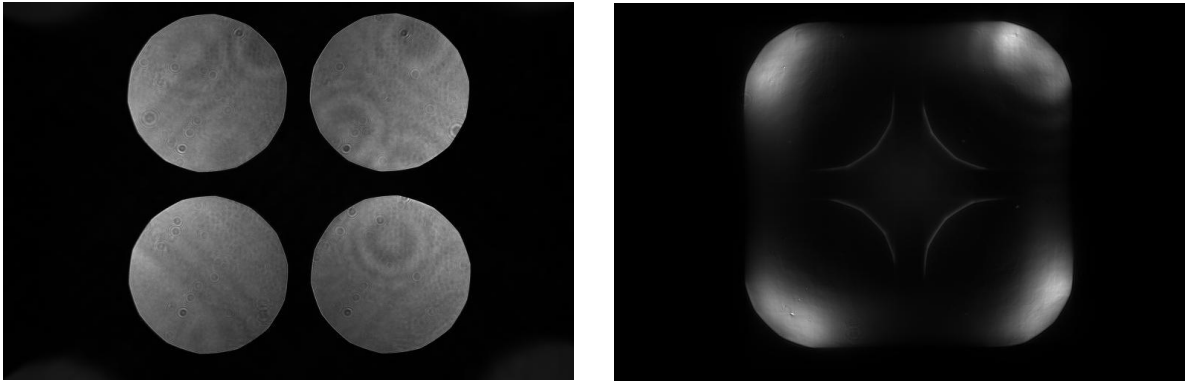

Figure 9. Left: 4 pupils image in the extreme point of the curve represented in Figure 8. Right: for comparison, an example of 4 pupils image showing the typical light distribution due to defocus WFE. In this case the WFS is saturated and out from its dynamical range.

**4.4 Phase Screens measurements**

As a last test, we measured the phase masks described in Section 3.1 to show the resolution capability of ASONG. The masks have been measured for reference by a Fizeau interferometer (4D Accufiz, 4 inch, 633 nm, stabilized, 9Mpx) with a 100 mm to 33 mm optical beam compactor to maximize the measurement resolution. Each of the phase measurements performed by the interferometer has therefore about a factor of 3 higher resolution w.r.t. ASONG existing prototype, but designs with higher resolution requirements can be customized. Table 2 shows the ASONG reconstructed WFE (left column) and the one measured by the Fizeau interferometer (right column). The ASONG WFE slopes are converted to WFE map by an iterative Fourier method[10]. This method computes the Fourier transform of the WF derivatives, divides it by $(u^2 + v^2)$ and makes an inverse Fourier transform, to retrieve the initial WF. The process is then iterated. When the error between two iterations becomes smaller than a preassigned level, the next WF estimate is computed and the iterative process is stopped.

This experimental test together with phase signal retrieval demonstrates the high spatial resolution capability of the ASONG concept (see below). Concerning, absolute comparison to interferometer measurements (of the order of tenths of lambda), the current experiment is actually limited by non-common path aberrations in between the two measurements set-up, rather than by the sensor itself. A new measurement campaign, including set-up aberration calibration and the addition of references in the phase screens to guide the matching algorithm, are foreseen for the next future (in order to extend our current estimations based on known defocus).

Table 2. Phase mask measurements

| | ASONG | Interferometer | WFE (µm) | |
| --- | --- | --- | --- | --- |
| | | | rms | PtV |
| **'Kolmo'** | 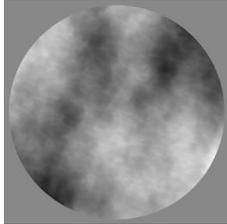 | 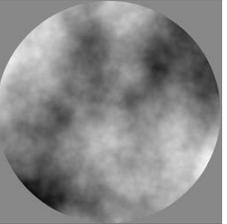 | 1.2 | 6.6 |
| **'Sinus'** | 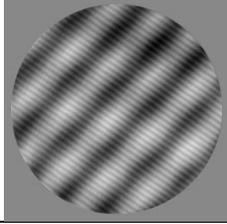 | 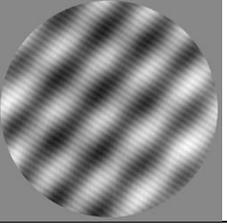 | 0.3 | 1.4 |
| **'Eye'** | 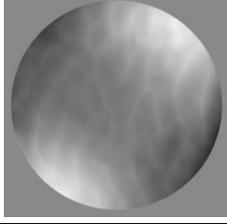 | 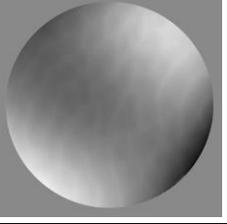 | 0.7 | 3.9 |
| **'Zer55'** | 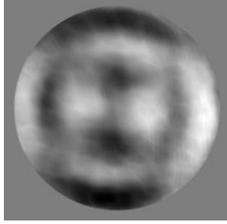 | 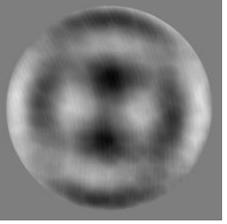 | 0.3 | 1.7 |
| **'Defocus'** | 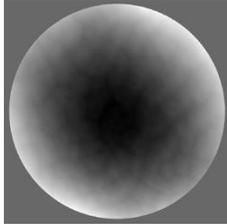 | 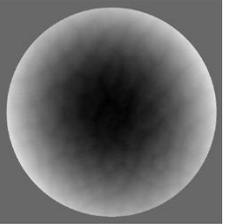 | 0.9 | 3.6 |
| **'Residual'** | 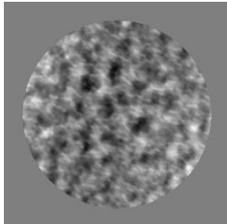 | 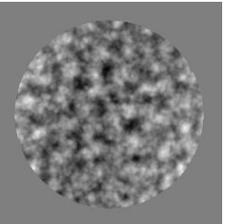 | 0.2 | 1.41 |

The capability of a low cost and compact device like ASONG to resolve very small WFE details already appears remarkable. A slight ripple pattern due probably to small residual phase introduced during the phase mask manufacturing while turning one surface of the polymer-polymer sandwich flat is clearly visible in all the phase masks measurements. This effect should add nominally a phase term of about 10 nm PtV.

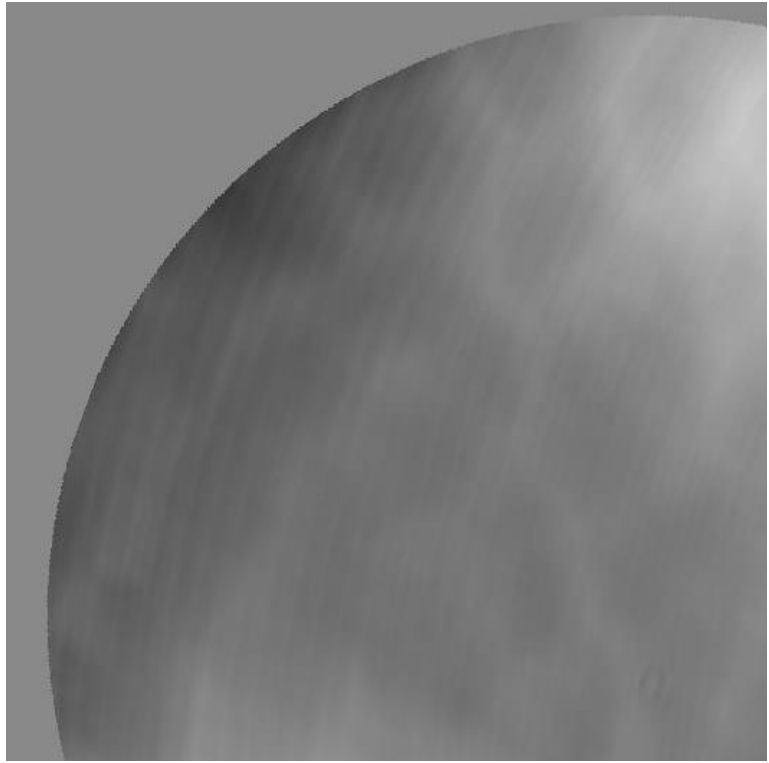

Figure 10. Zoomed detail of the 'eye' phase mask measurement. A few very high order 'defects' in mask manufacturing are clearly visible.

## 5. CONCLUSIONS AND FUTURE DEVELOPMENTS

We have shown in this paper the first results of the characterization of first prototype of the ASONG WFS. This WFS collects in an elegant and simple solution, high spatial resolution, high sensitivity and a wide dynamical range. These characteristics make it particularly interesting for different applications, from ophthalmology, to optical metrology and possibly active/adaptive optics. This last application has not been studied yet in terms of capability to work in low SNR conditions, but the sensor might find an application in small telescopes aberration correction. Different use cases will be studied will be studied in the near future.

## AKNOWLEDGEMENTS

This work is funded by Linksium maturation project (# 200032) and was previously supported by CNRS (Centre National de la Recherche Scientifique) innovation project from 2019 to 2021. We gratefully thank Linksium for topical and intellectual discussions about the research and consideration on technology transfer from academia to industry.

# REFERENCES

[1] J. W. Hardy, J. E. Lefebvre, and C. Koliopoulos, "Real-time atmospheric compensation," JOSA, 67(3), 360-369, (1977).
[2] T. O. Salmon, L. N. Thibos, and A. Bradley, "Comparison of the eye's wave-front aberration measured psychophysically and with the Shack–Hartmann wave-front sensor," JOSA A, 15(9), 2457-2465, (1998).
[3] C. R. Forest, C. R. Canizares, D. R. Neal et al., "Metrology of thin transparent optics using Shack-Hartmann wavefront sensing," Optical engineering, 43(3), 742-754, (2004).
[4] Y. Feng, F. Hénault, L. Schreiber, A. Spang, "Development and implementation of crossed-sine wavefront sensor for simultaneous high spatial resolution imaging," Proceedings of the SPIE, Volume 11451, id. 1145145 12 pp. (2020).
[5] F. Hénault, A. Spang, Y. Feng et al., "Crossed-sine wavefront sensor for adaptive optics, metrology and ophthalmology applications," Engineering Research Express, 2(015042), (2020).
[6] F. Henault, A. Spang, "System for inspecting surfaces of an optical wave using a graduated density filter", WO2020156867 (2020).
[7] F. Hénault, "Fresnel diffraction analysis of Ronchi and reverse Hartmann tests," JOSA A, 35(10), 1717-1729, (2018).
[8] J. Porter, A. Guirao, I. G. Cox, D. R. Williams "Monochromatic aberrations of the human eye in a large population," J Opt Soc Am A Opt Image Sci Vis, Volume 18, No 8 (2001).
[9] G. Herriot, P. Hickson, B. Ellerbroek, J-P. Véran, C-Y. She, R Clare, D. Looz "Focus errors from tracking sodium layer altitude variations with laser guide star adaptive optics for the Thirty Meter Telescope," Advances in Adaptive Optics II, Proc. of SPIE Vol. 6272, 62721I, (2006).
[10] F. Roddier, C. Roddier "Wavefront reconstruction using iterative Fourier transforms," Applied Optics, 30, 1325, (1991).